\documentstyle[12pt, a4]{article}
\begin{document}
\author{B. G\"{o}n\"{u}l
\and Department of Engineering Physics, University of Gaziantep, 
\and 27310 Gaziantep-T\"{u}rkiye}
\title{Generalized Supersymmetric Perturbation Theory}
\date{}
\maketitle

\begin{abstract}
Using the basic ingredient of supersymmetry, we develop 
a simple alternative approach to perturbation theory in one-dimensional 
non-relativistic quantum mechanics. The formulae for the energy 
shifts and wave functions do not involve tedious calculations 
which appear in the available perturbation theories. The model 
applicable in the same form to both the ground state and excited 
bound states, unlike the recently introduced supersymmetric perturbation 
technique which, together with other approaches based on logarithmic 
perturbation theory, are involved within the more general framework 
of the present formalism.
\end{abstract}

\section{Introduction}
An exact solution of the Schr\"{o}dinger equation exists only for 
a few idealized problems; normally it has to be solved using 
an approximation method such as the perturbation theory (PT), 
which constitutes one of the most powerful tools available in 
the study of quantum mechanics in the atoms and molecules. PT 
is applied to those cases in which the real system can be described 
by a small change in an exactly solvable idealized system. In 
this form we can describe a great number of problems encountered 
especially in atomic physics, in which the nucleus provides the 
strong central potential for the electrons; further interactions 
of less strength are described by the perturbation. Examples 
of these additional interactions are: the magnetic interaction 
(spin-orbit coupling), the electrostatic repulsion of electrons 
and the influence of external fields. But in spite of widespread 
application of this theory, its basic analytical properties are 
poorly understood. One of our objectives in this paper is to 
illustrate selected important aspects of the perturbation theory 
within the frame of supersymmetric quantum mechanics.

Performing explicit calculations in non-relativistic quantum 
mechanics using the familiar Rayleigh-Schr\"{o}dinger perturbation 
expansion is rendered difficult by the presence of summations 
over all intermediate unperturbed eigenstates. Alternative perturbation 
procedures have been proposed to avoid this difficulty, notably 
the logaritmic perturbation theory (LPT) \cite{aharonov15821979}-\cite{au38371991} 
and the Dalgarno-Lewis 
technique \cite{dalgarno2331955}-\cite{mavromatis7381991}. The virtue 
of LPT is its avoidance of the cumbersome 
summation over states for second- and higher-order corrections 
in Rayleigh-Schr\"{o}dinger perturbation theory. Unfortunately, 
it has problems of its own in calculating corrections to excited 
states, owing to presence of nodes in the wave functions. Various 
schemes have been proposed to circumvent the resulting singularities 
\cite{au38371991, kim251992, dobrovolska5631999}.

Such is the status of LPT after over 20 years of active development. 
Meanwhile, supersymmetric quantum mechanics 
(SSQM) \cite{cooper2671995, junker1996} has 
developed immensely since the first models were 
introduced \cite{nicolai14971976, witten5131981}. 
Several approximation methods using SSQM formalism have been 
developed, including the supersymmetric perturbation theory (SSPT) 
of Cooper and Roy \cite{cooper2021990}. Recently, 
Lee \cite{lee1012000} has shown that SSPT 
and LPT are entirely equivalent and fortuitously, each turns 
out to resolve difficulties encountered in the other. Namely, 
LPT formulas for energy corrections obviate tedious procedures 
in the SSQM method, while the use of SSQM partner potentials 
with virtually identical bound state spectra solves difficulties 
with excited states encountered in LPT. Although the iterative 
procedure in SSPT may not actually reduce the calculational workload, 
it does cast the calculations into a physically-motivated, visualizable 
framework.

In this letter, starting from the first priciples, we develop 
a more economical scheme which yields simple but closed perturbation 
theory formulae leading to the Riccati equation from which one 
can actually obtain all the perturbation corrections to both 
energy level shifts and wave functions for all states, unlike 
the other models mentioned above. In the application of the present 
method to the $n^{th}$ excited state, one requires knowledge of 
the unperturbed eigenfunction $\chi_{n} (r)$ but 
no knowledge of the other eigenvalues or eigenfunctions 
is necessary. The procedure introduced here does not involve 
either tedious explicit factoring out of the zeros of 
$\chi_{n} (r)$ \cite{aharonov15821979, au22451979} or 
introduction of ghost states \cite{au38371991} as were the cases 
encountered for applying LPT to excited states. Since, the present 
model offers the explicit expressions for the energy corrections, 
which are absent in the original SSPT while the treatment of 
Lee \cite{lee1012000} for such calculations has mathematical complexity, and 
provides a clean route to the excited states, which are cumbersome 
to analyze in both LPT and SSPT, our results can be thought of 
as a generalization of logarithmic and supersymmetric based perturbation 
theories. This is the another objective in the present work.

In the following section we introduce the model and discuss briefly 
the phyiscs behind the formulation. In Section 3, some applications 
are given and the power of the present technique is illustrated 
when compared to the calculation technique of other theories 
considering the whole of states. Some concluding remarks and 
summary of the work are drawn in the last section. 

\section{Formulation}
The goal in SSQM is to solve the Riccati equation,
\begin{equation}
W^{2}(r)-\frac{\hbar}{\sqrt{2m}}W'(r)=V(r)-E_{0}~,
\end{equation}
where $V(r)$ is the potential of interest and $E_{0}$ is 
the corresponding ground state energy. If we find  $W(r)$, 
the so called superpotential, we have of course found the ground 
state wave function via,
\begin{equation}
\psi_{0}(r)=N~\exp\left[-\frac{\sqrt{2m}}{\hbar}\int_{}^{r}W(z)dz\right]~,
\end{equation}
where $N$ is the normalization constant. If $V(r)$ is a 
shape invariant potential, we can in fact obtain the entire 
spectrum of bound state energies and wave functions via ladder 
operators \cite{cooper2671995}. Through the article, this basic ingredient of 
SSQM given by (1) and (2) will be extended and used for the treatment of 
excited states.

Now, suppose that we are interested in a potential for which 
we do not know $W(r)$ exactly, and the corresponding Hamiltonian is not factorizable 
but almost factorizable. More specifically, we assume that $V(r)$ differs 
by a small amount from a potential 
$V_{0} (r)$ plus angular momentum barrier if any, for which one solves the 
Riccati equation explicitly. For the consideration of spherically 
symmetric potentials, the corresponding Schr\"{o}dinger equation 
for the radial wave function has the form 
\begin{equation}
\frac{\hbar^{2}}{2m}\frac{{\psi}''_{n}(r)}{\psi_{n}(r)}=\left[V(r)-E_{n}\right]
~~,~~
V(r)=\left[V_{0}(r)+\frac{\hbar^{2}}{2m}\frac{\ell(\ell+1)}{r^{2}}\right]+\Delta V(r),
\end{equation}
where $\Delta V$ is a perturbing potential. Let us write the wave function 
$\psi_{n}$ as
\begin{equation}
\psi_{n}(r)=\chi_{n}(r)\phi_{n}(r)~,
\end{equation}
in which 
$\chi_{n}$ is the known normalized eigenfunction of the unperturbed Schr\"{o}dinger 
equation whereas $\phi_{n}$ is a moderating function corresponding to the perturbing potential. 
Substituting (4) into (3) yields 
\begin{equation}
\frac{\hbar^{2}}{2m} \left( \frac{{\chi}''_{n}}{\chi_{n}}
+\frac{{\phi}''_{n}}{\phi_{n}} +2\frac{\chi '_{n}}{\chi_{n}}
\frac{\phi'_{n}}{\phi_{n}} \right) =V-E_{n}~.
\end{equation}
Instead of setting the functions 
$\chi_{n}$ and $\phi_{n}$, we will set their logarithmic derivatives using the spirit 
of Eqs. (1) and (2);
\begin{equation}
W_{n} =-\frac{\hbar }{\sqrt{2m}} \frac{\chi '_{n}}{\chi_{n}}
~~,~~
\Delta W_{n} =-\frac{\hbar }{\sqrt{2m}} \frac{\phi '_{n}}{\phi_{n}}
\end{equation}
which leads to 
\begin{equation}
\frac{\hbar^{2}}{2m} \frac{{\chi}''_{n}}{\chi_{n}} =W_{n}^{2}
-\frac{\hbar }{\sqrt{2m}} W'_{n} =\left[ V_{0} (r)+\frac{\hbar^{2}
}{2m} \frac{\ell (\ell +1)}{r^{2}} \right] -\epsilon _{n}~,
\end{equation}
where $\epsilon _{n}$ is the eigenvalue of the unperturbed and exactly solvable potential, 
and 
\begin{equation}
\frac{\hbar^{2}}{2m} \left( \frac{{\phi}''_{n}}{\phi_{n}}
+2\frac{\chi '_{n}}{\chi_{n}} \frac{\phi '_{n}}{\phi_{n}} \right)
=\Delta W_{n}^{2}-\frac{\hbar}{\sqrt{2m}}\Delta W'_{n}+2W_{n}
\Delta W_{n} =\Delta V(r)-\Delta \epsilon _{n}~,
\end{equation}
in which $\Delta \epsilon _{n}$ is the eigenvalue for the perturbed potential, and 
$E_{n} =\epsilon _{n} +\Delta \epsilon _{n}$. 
Then, Eq. (5), and subsequently Eq. (3), reduces to 
\begin{equation}
\left( W_{n} +\Delta W_{n} \right) ^{2} -\frac{\hbar }{\sqrt{2m}} \left(
W_{n} +\Delta W_{n} \right)' =V-E_{n}~,
\end{equation}
which is similar to Eq. (1). In priciple as one knows explicitly 
the solution of Eq. (7), namely the whole spectrum and corresponding 
eigenfunctions of the unperturbed interaction potential, the 
goal here is to solve only Eq. (8), which is the main result 
of this letter, leading to the solution of Eqs. (3) and (9). 

Eq. (8) is a closed analytical form in comparing to lengthy SSPT 
and LPT expressions, in particular for the excited states. In 
this respect, the present formulation has a more general form 
than the available perturbation theories. Though this point will 
be clear in the next section through the applications, it would 
be convenient at this stage to clarify how Eq. (8) involves in 
a compact form the supersymmetric and logarithmic perturbation 
theory expressions. As the equivalance of SSPT to LPT has already 
been clarifed \cite{lee1012000}, we consider here only the framework of SSPT 
and show that SSPT is a subset of the present model.

For the perturbation technique, we have initially assumed that 
we could split the given potential in two parts, Eq. (3). The 
main part corresponds to a shape invariant potential, Eq. (7), 
for which the superpotential is known analytically and the remaining 
part is treated as a perturbation, Eq. (8). If necessary, one 
can expand the functions related to the perturbation in terms 
of the perturbation parameter $\lambda $,
\begin{eqnarray}
\Delta V(r;\lambda ) & = & \sum\limits_{k=1}^{\infty}\lambda^{k}\Delta V_{k}(r)~,
\nonumber
\\
\Delta W_{n}(r;\lambda)& = & \sum\limits_{k=1}^{\infty}\lambda^{k}\Delta W_{nk}(r)~,
\nonumber
\\
\Delta\epsilon_{n}(\lambda) & = & \sum\limits_{k=1}^{\infty}\lambda^{k}\epsilon_{nk}~,
\end{eqnarray}
where $\lambda $ will eventually be set equal to one. Substitution of the above 
expansion into Eq. (8) by equating terms with the same power of $\lambda $
 on both sides yields up to $O\left( \lambda ^{3} \right) $
\begin{equation}
2W_{n} \Delta W_{n1} -\frac{\hbar }{\sqrt{2m}} \Delta W'_{n1} =\Delta
V_{1} -\Delta \epsilon _{n1}~,
\end{equation}
\begin{equation}
\Delta W_{n1}^{2} +2W_{n} \Delta W_{n2} -\frac{\hbar }{\sqrt{2m}} \Delta
W'_{n2} =\Delta V_{2} -\Delta \epsilon _{n2}~,
\end{equation}
\begin{equation}
2\left( W_{n} \Delta W_{n3} +\Delta W_{n1} \Delta W_{n2} \right)
-\frac{\hbar }{\sqrt{2m}} \Delta W'_{n3} =\Delta V_{3} -\Delta \epsilon
_{n3}~,
\end{equation}
which are exactly SSPT expressions appeared 
in \cite{cooper2021990}-\cite{chakrabarti2852001} but for 
the case $n=0$. Eq. (8) and its expansion, Eqs. (11-13), give a flexibility 
for the easy calculations of the perturbative corrections to 
energy and wave functions for the $n^{th}$ state of interest 
through an appropriately chosen perturbed 
superpotential, unlike the other perturbation theories. We will 
show in the next section through the applications that this feature 
of the present model leads to a simple framework in obtaining 
the corrections to all states without using complicated and tedious 
mathematical procedures. 

\section{Application}
\subsection{perturbed harmonic oscillator}
Let us start with the most elementary kind of perturbation calculation. 
Consider a perturbed harmonic oscillator potential of the form
\begin{equation}
V(r;\lambda )=\left[ \frac{1}{2} mw^{2} r^{2} +\frac{\ell \left( \ell
+1\right) \hbar^{2}}{2mr^{2}} \right] +\frac{1}{2} m\lambda w^{2} r^{2}~,
\end{equation}
which woud arise by increasing the spring constant of a harmonic 
oscillator from $K$ to $(1+\lambda )K$ since $w=\sqrt{K/m }$. The 
perturbed potential in its present form can of course be 
solved exactly. Neverthless, we want to solve this problem in 
the light of our approach to test the effectiveness of the present 
technique. This application will also clarify that our derivation 
is much simpler and more direct than the other methods, and hence 
provides a useful alternative.

The whole spectrum and corresponding wave functions for the unperturbed 
part of the problem is well known in the literature. Then, starting 
with the normalized wave function of the harmonic oscillator system 
for any state of interest, one can easily calculate the corrections 
at one step to the energy and eigenvalues for the perturbed potential 
setting an appropriate superpotential satisfying Eq. (8). For 
the comparison of our results with those of recent works, we 
will perform the calculations here for the ground state $n=0$. 
Starting with Eq. (8),
\begin{equation}
\Delta W_{n=0}^{2}-\frac{\hbar }{\sqrt{2m}} \Delta W'_{n=0} +2W_{n=0}
\Delta W_{n=0} =\frac{1}{2} m\lambda w^{2} r^{2} -\Delta \epsilon _{n=0}~,
\end{equation}
one readily sees that the perturbed superpotential ($\Delta W$) 
satisfying the above equation has the form
\begin{equation}
\Delta W_{n=0} =\sqrt{\frac{m}{2}} wr\left( \sqrt{1+\lambda } -1\right)~,
\end{equation}
since, from the literature \cite{cooper2671995}, the superpotential $W$ 
corresponding to the unperturbed potential is in the form 
\begin{equation}
W_{n=0} =\sqrt{\frac{m}{2}} wr-\frac{\left( \ell +1\right) \hbar
}{\sqrt{2m} r}~.
\end{equation}
Upon substituting (16) and (17) into (15) yields
\begin{equation}
\Delta \epsilon _{n=0} =\left( \ell +\frac{3}{2} \right) ~\left(
\sqrt{1+\lambda } -1\right) ~\hbar w~,
\end{equation}
and using (16) with (2) leads to the moderating function for 
the ground state 
\begin{equation}
\phi_{n=0} =\exp \left[ \frac{mw}{2\hbar } r^{2} \left(
1-\sqrt{1+\lambda } \right) \right]~.
\end{equation}
As the ground state wave function $\chi_{n=0}$ corresponding 
to the superpotential in (17) and energy $\epsilon _{n=0}$ 
for the normalized unperturbed harmonic oscillator are 
\begin{equation}
\chi_{n=0}=N~r^{\ell +1}\exp \left( -\frac{mwr^{2}}{2\hbar }
\right) ~~,~~\epsilon _{n=0} =\left( \ell +\frac{3}{2} \right)
\hbar w~,
\end{equation}
the total wave function $\psi _{n=0}$ for the whole system in (14) and corresponding energy are obtained 
as
\begin{equation}
\psi _{n=0}=\chi_{n=0} \phi_{n=0}=N~r^{\ell +1}\exp \left(
-\frac{mw\sqrt{1+\lambda }}{2\hbar } r^{2} \right)
~,~
E_{n=0}=\sqrt{1+\lambda}~\left(\ell +\frac{3}{2}\right)~\hbar w~.
\end{equation}
These are indeed the correct results which can be verified explicitly 
by the use of Eq. (9), since this problem has an exact analytical 
solution.

Lee \cite{lee1012000,lee1999} recently has studied this problem within the frame 
of supersymmetric perturbation theory and worked out it up to the 
second order. From the expansion of the closed expressions in 
(21) in $\lambda $ such that 
$\sqrt{1+\lambda } =1+\lambda /2 -\lambda ^{2} /8 +\ldots $, 
it can clearly be seen that Lee's treatment indeed appears 
as a subset of the present calculation procedure. In addition, 
our sophisticated approach neither involves cumbersome procedures 
nor tedious calculations. In what follows, we further show that 
the present model is also applicable in the same form to bound 
excited states without any difficulty, unlike the works 
in \cite{cooper2021990}-\cite{lee1999}.

As the potential considered in (14) is a shape invariant potential, 
using (9) together with (16) and (17) for $n=0$ and having in mind the shape invariance property for the exactly 
solvable potentials \cite{cooper2671995},
\begin{equation}
V_{+} (r,a_{0} )=V_{-} (r,a_{1} )+R(a_{1} )~~;~~R\left( a_{1}
\right) =2\hbar w\sqrt{1+\lambda }~,
\end{equation}
where $V_{-}$ is the exactly solvable potential in (9) 
and $V_{+}$ is its supersymmetric partner 
\begin{equation}
\left( W_{n=0} +\Delta W_{n=0} \right) ^{2} +\frac{\hbar }{\sqrt{2m}}
\left( W_{n=0} +\Delta W_{n=0} \right)' =V_{+} -E_{n=0}~,
\end{equation}
and $a_{0} =\ell $, $a_{1} =\ell +1$. Therefore, 
\begin{equation}
E_{n} =\sum\limits_{s=1}^{n}R(a_{s} )+\left( \ell +\frac{3}{2} \right) 
\hbar w=\left( 2n+\ell +\frac{3}{2} \right) \hbar w\sqrt{1+\lambda }~,
\end{equation}
which are the whole spectrum for the perturbed harmonic potential 
in (14). This very simple procedure removes difficulties encountered 
in SSPT and LPT in dealing with excited states. Again, this result 
agrees with Eq. (47) of \cite{lee1012000} in which the excited state energies 
are presented up to only second order due to nasty calculation 
procedure.

To find the energy of the excited states, all we had to do was 
to perform calculations using a proper superpotential corresponding 
to the ground state wave function of the perturbed potential. 
The LPT literature goes to great pains to find ways to avoid 
using excited state wave functions. The study of present example, 
in particular the procedure used through Eqs. (22) to (24), thus 
illustrates in this respect how SSQM allows us to use the LPT 
based formulas for the entire spectrum, by moving to a partner 
potential whose ground state coincides with the excited state 
of the potential initillay considered. The reader is referred 
to \cite{cooper2671995} for a further discussion on supersymmetric partner 
potentials and their applications.

\subsection{perturbed Coulomb interaction}
Kim and Sukhatme \cite{kim251992} obtained a set of expressions for ground 
and excited state wave functions and energies in perturbation 
theory (henceforth referred to as KS), that do not involve infinite 
sums, and which they consider a generalization of LPT. Subsequently 
some expressions in this formalism were simplified by Mavromatis 
\cite{mavromatisL5151993}. Later, KS approach were carefully 
compared \cite{mavromatis12021996} with the 
results of the LPT formalism and step by step connection between 
the two formalisms was shown and commented on via an illustrative 
example, together with the discussion on the relation between 
KS approach to Dalgarno-Lewis 
formalism \cite{dalgarno2331955}-\cite{mavromatis7381991}.

As we are aware of the relation between LPT and SSPT, one then 
may expect also the connection between SSPT and LPT based KS 
approach. Thus, we focus through this section and the other examples 
in the next sections on constructing a bridge between the present 
generalized supersymmetric perturbation formalism and KS approach 
and show explicitly the equivalence between the two aproaches, 
considering of course the expanded form of Eq. (8). 

We proceed first with the present treatment of the simple example 
used in \cite{mavromatis12021996}, which is the perturbed Coulomb system,
\begin{equation}
V(r)=\left[ -\frac{e^{2}}{r} +\frac{\ell \left( \ell +1\right) \hbar
^{2}}{2mr^{2}} \right] +\lambda \frac{e^{2}}{2r}~,
\end{equation}
where, for a particular example, $\lambda $ denotes the increse in the charge of the nucleus in case of 
an interaction between electron and nucleus in hydrogen like atoms.

The superpotential, wave function and energy for the ground state 
corresponding to the unperturbed potential, respectively, are \cite{cooper2671995}
\begin{eqnarray}
W_{n=0} & = & \sqrt{\frac{m}{2}}\frac{e^{2}}{\left(\ell +1\right)\hbar}-
\frac{\left(\ell +1\right)\hbar}{\sqrt{2m}r}~,
\nonumber
\\
\chi_{n=0} & = & N~r^{\ell +1}\exp\left(-\frac{me^{2}}{\left(\ell +1\right)\hbar^{2}}r\right)~,
\nonumber
\\
\epsilon_{n=0} & = & -\frac{me^{4}}{2\hbar^{2}\left(\ell +1\right)^{2}}~.
\end{eqnarray}
From Eq. (8), $\Delta W_{n=0}$ involving all the corrections corresponding to the perturbing 
potential $\Delta V=\lambda e^{2} /2r $ is readily obtained as
\begin{equation}
\Delta W_{n=0}=-\sqrt{\frac{m}{8}}\frac{\lambda e^{2}}{\left( \ell +1\right)\hbar}~,
\end{equation}
and the the total correction to the enery $\Delta \epsilon _{n=0}$, 
together with the moderating function due to the perturbing potential are
\begin{equation}
\Delta \epsilon _{n=0} =-\frac{me^{4}}{2\hbar^{2}} \left(
\frac{\lambda ^{2}}{4} -\lambda \right) ~~,~~\phi_{n=0} =\exp
\left( \frac{\lambda me^{2}}{2\hbar \left( \ell +1\right) } r\right)~.
\end{equation}
Hence, the full ground state wave function and energy spectrum 
for the perturbed Coulomb system in (25) are
\begin{eqnarray}
\psi_{n=0}=\chi_{n=0} \phi_{n=0} & = & N~r^{\ell +1} \exp \left[
-\frac{me^{2}}{\left( \ell +1\right) \hbar^{2}} \left( 1-\frac{\lambda
}{2} \right)~r \right] ~,
\nonumber
\\
E_{n=0} & = & -\frac{me^{4}}{2\hbar^{2}
\left( \ell +1\right) ^{2}} \left( 1-\frac{\lambda }{2} \right) ^{2}~.
\end{eqnarray}
One can justify the result above via Eq. (9) and the full excited 
energy spectrum can be easily found using the procedure given 
through Eqs (22-24), for which the superpotential should have 
the final form $W_{n=0} +\Delta W_{n=0}$.

Now, to see the close relation between the formalism presented 
in this letter and that of Kim and Sukhatme \cite{kim251992}, the reader should 
go back to Eqs. (11-13) which are the expansion of the present 
formalism yielding individual corrections in order, having in 
mind that $\phi_{n}^{KS}$ is expanded in orders of $\lambda $ in 
KS approach. In the light of recent works \cite{lee1012000, lee1999} regarding 
SSPT, one sees after some algebra that the expansion terms of 
moderating function $\phi_{n}^{KS}$ in KS technique 
are related to those of the superpotential $\Delta W_{n}$ 
appearing in (11) through (13). For 
clarity consider only the first order, then 
\begin{equation}
\Delta W_{n1} =-\frac{\hbar }{\sqrt{2m}} \frac{d}{dr} \phi_{n1\quad
}^{KS} ~~,~~
\end{equation}
since, from (11), 
\begin{equation}
\Delta W_{n1} (r)=\frac{\sqrt{2m}}{\hbar } \frac{1}{\chi_{n}^{2} (r)}
\int\limits_{-\infty }^{r}\chi_{n}^{2} (z)\left[ \Delta \epsilon _{n1}
-\Delta V_{1} (z)\right] dz~.
\end{equation}
Mavromatis \cite{mavromatis12021996}, using KS and LPT formalism and working within $\hbar =m=e^{2} =1$ 
unit system for $\ell =0$ case, investigated the same problem for the ground state solution 
$\left( n=0\right) $ for which after some tedious integrals he arrived at
\begin{equation}
\phi_{01}^{KS} =\frac{r-a}{2} \lambda ~~,~~\phi_{02}^{KS}
=\frac{\left( r-a\right) ^{2}}{8} \lambda ^{2} ~~,~~\phi
_{03}^{KS} =\frac{\left( r-a\right) ^{3}}{48~~} \lambda ^{3}~~,\ldots
\end{equation}
and $\Delta \epsilon _{01} =\lambda /2 $, 
$\Delta \epsilon _{02} =-\lambda ^{2} /8 $ 
with higher order energy corrections being zero. We should remark 
at this point that the present approach has provided exact result 
in a closed analytical form, Eq. (28), without dealing with nasty 
integrals, since the moderating function in (32) eventually can 
be written as 
$\phi_{n}^{KS} =\exp \left[ \lambda /2\left( r-a\right)  \right] $ with 
an extra constant, $\exp \left( -\lambda a/2 \right) $, though 
the constants do not affect the energy, where $a$ comes 
from the lower limit of the integrals carried out. Considering 
Eqs. (30,32) within the same frame and the whole discussion given 
in this section, together with the physics behind Eqs. (8-13), 
one can see that the model introduced here through Eq. (8) unifies 
not only SSPT and LPT but also the KS approach. In the following 
section, we will make more clear this point with the further 
examples used in \cite{kim251992}. 

\subsection{harmonic oscillator with linear perturbation (n=1 state)}
Kim and his co-worker considered a one dimensional perturbed 
harmonic oscillator potential with no angular momentum barrier 
\begin{equation}
V=\frac{1}{4} w^{2} r^{2} +\lambda r+B~,
\end{equation}
involving a linear perturbation term. Picking the particular 
case of an applied uniform electric field to a charged particle 
moving in a simple harmonic oscillator potential provides a simple 
physical interpretation to the linear term in (33) where in this 
case $\lambda $ stands for the strenght of the field applied. This explanation 
is also valid for the next example considered in the following 
section but for a different unperturbed potential.

In \cite{kim251992}, the energy level shifts and corrections to the wave 
function were worked out up to the second order for the first 
excited state $\left( n=1\right) $,
\begin{equation}
\Delta \epsilon _{11} =B~~,~~\Delta \epsilon _{12}
=-\frac{\lambda ^{2}}{w^{2}} ~~,~~~~\phi_{11}^{KS}
=-\frac{\lambda }{w} \left( r-\frac{2}{wr} \right) ~~,~~\phi
_{12}^{KS} =\frac{\lambda ^{2} r^{2}}{2w^{2}}~.
\end{equation}
For a clear comparison we work within $\hbar =2m=1$ unit 
system as in \cite{kim251992}. From (30) and (34), $\Delta W$ in the first order should read
\begin{equation}
\Delta W_{11} =\frac{\lambda }{w} \left( 1+\frac{2}{wr^{2}} \right)~,
\end{equation}
and the use of (35) in (11) yields explicitly the first order 
energy shift $\Delta \epsilon _{11} =B$, since 
\begin{equation}
\chi_{n=1} =N~r~\exp \left( -wr^{2} /4 \right) ~~,~~W_{n=1}
=\frac{wr}{2} -\frac{1}{r} ~~,~~\epsilon _{n=1} =3w/2~~.
\end{equation}
The second order calculations can be carried out in the same 
manner for comparison. However, our aim here is to clarify the 
effectiveness of the technique introduced in this article by 
calculating all corrections explicitly in a simple way. For this 
reason we turn back our attention to the shifted harmonic oscillator 
potential in (33) in order to remind the reader that the potential 
in (33) can be exactly solved in its present form by setting 
the superpotential 
\begin{equation}
W_{n=1}^{exact} =\frac{wr}{2} +\frac{\lambda }{w}~,
\end{equation}
from which one sees that $B=-w$ and 
$E_{n=1}^{exact} =3w/2 +B-\lambda ^{2} /w^{2}  =\epsilon _{n=1} +\Delta
\epsilon _{11} +\Delta \epsilon _{12}$ which justifies the 
results obtained above. This makes clear that the energy 
level shifts due to higher orders in $\lambda $ vanish. Eq. (37) leads 
to the full wave function in the light of Eq. (6)
\begin{equation}
\psi _{n=1}^{exact} =N_{1} ~\exp \left( -wr^{2} /4 \right) \exp \left(
-\lambda r/w \right)~.
\end{equation}
Further, it is not difficult to see that Eq. (9) provides the 
exact full energy spectrum involving all excited states using 
the shape invariance property of the potential in (33) without 
carrying out tedious integrals as in the other perturbation theories. 
The physics behind this exact solution would shed a light to 
be able to use Eq. (8) for an easy calculation of the contributions 
comes from due to the perturbed terms, which are briefly discussed 
below.

To illustrate once more the power and elegancy of the present 
formalism we here show that all the corrections to the sytem of interest 
can be calculated simply by Eq. (8). For the choice of correct 
superpotential leading to the linear perturbation term we use 
\begin{equation}
\Delta W_{n=1} =W_{n=1}^{exact} -W_{n=1} =\frac{\lambda }{w} +\frac{1}{r}~,
\end{equation}
and upon substituting (39) into (8) one clearly sees that full 
energy corrections are 
$\Delta \epsilon _{n=1} =\Delta \epsilon _{11} +\Delta \epsilon _{12}
=B-\lambda ^{2} /w^{2}$, together with the corresponding moderator function involving 
all the modifications
\begin{equation}
\phi_{n=1} =\frac{\exp \left( -\lambda r/w \right) }{r}~.
\end{equation}
Hence, the first excited state wave function for (33) reads
\begin{equation}
\psi _{n=1} =\chi_{n=1} \phi_{n=1} =N~\exp \left( -wr^{2} /4 \right)
\exp \left( -\lambda r/w \right)~,
\end{equation}
from which it is clear that the present formalism produces exact 
results. However, the correction terms in KS approach \cite{kim251992} obtained 
for $n=1$ state up to the second order given in (34) deviate slightly 
from the exact result (see Eq. (25) of \cite{kim251992}) as in the previous 
example, while it leads to the correct answer for the correction 
terms to the energy.

\subsection{infinite square well with linear perturbation (n=2 state)}
Finally, Kim and Sukhatme used their technique to investigate 
the $n=2$ state of an infinite square well potential
\begin{equation}
V(r)=
\left
\{
\begin{array}{ll}
0 & if |r|\leq\pi/2
\\
\infty & otherwise
\end{array}
\right.~,
\end{equation}
subject to a perturbation $\Delta V(r)=\lambda r+B$. To 
calculate the energy level shifts and corrections to 
the unperturbed wave function
\begin{equation}
\chi_{n=2} =\sqrt{\frac{2}{\pi }} \cos 3r~,
\end{equation}
one first needs to define the corresponding superpotential
\begin{equation}
W_{n=2} =-\frac{\chi '_{n=2}}{\chi_{n=2}} =3\tan 3r~.
\end{equation}
To see once more the close relation between KS approach and the 
present formalism, consider Eq. (11) in which $\Delta W_{21}$ can 
be obtained through (30)
\begin{equation}
\Delta W_{21} =\frac{\lambda \sec ^{2} 3r}{4} \left( \frac{\pi ^{2}}{4}
-r^{2} \right) -\frac{\lambda }{6} \left( r\tan 3r+\frac{1}{6} \right)~,
\end{equation}
since the second order correction to the wave function $\phi_{21}^{KS}$ in 
\cite{kim251992} appears in the form
\begin{equation}
\phi_{21}^{KS} =\frac{\lambda }{4} \left[ \frac{1}{3} \left( r^{2}
-\frac{\pi ^{2}}{4} \right) \tan 3r+\frac{1}{9} r\right]~.
\end{equation}
Eq. (45) can also be checked out using (31) by setting the 
limits of the integral from $-\pi /2 $ to $\pi /2 $. Substitution 
of Eqs. (44) and (45) into (11) verifies the result $\Delta \epsilon _{21} =B$ 
in \cite{kim251992}. Similarly, the second order 
energy level correction found in \cite{kim251992} is
\begin{equation}
\Delta \epsilon _{22} =\frac{\lambda ^{2}}{36} \left( \frac{\pi ^{2}
}{12} -\frac{5}{36} \right)~,
\end{equation}
which can in our theoretical framework be obtained by
\begin{equation}
\Delta \epsilon _{22} =-\int\limits_{-\pi /2 }^{\pi /2 }\frac{2}{\pi } 
\cos ^{2} 3r~\Delta W_{21}^{2} \left( r\right) dr~,
\end{equation}
that is derived from (12).

Again, the study of this example justifies that the formalism 
introduced in this letter for perturbative problems has a more 
general form, since the whole KS approach for the calculation 
of corrections due to perturbative terms is hidden in the subset 
of our formalism, namely the expansion of Eq. (8) leading to 
Eqs. (11) and (12) for the first and second order modifications 
reproduces same results as in KS formalism. Additionally, the 
mathematical treatment is much simpler in the frame of the present 
approach. 

\section{Concluding Remarks}
No single approximation method available in the literature 
is ideal for every problem. SSPT and LPT based theories 
avoids the Rayleigh-Schr\"{o}dinger summation, but it can lead 
to nasty integrals and more effort in particular for excited 
states. The method is valuable when the integrals can be done 
exactly or by a reliable numerical procedure. Otherwise, the 
Rayleigh-Schr{\"o}dinger summation, even when it does not give an 
exact answer, starts not to look so bad after all. This 
was the motivation behind the work introduced in this 
article. The present perturbation model appears in this 
respect to be superior for the excited states and 
provides a quick route to the calculation of all 
corrections within the frame of the perturbation 
theory, which considerably simplify one's 
calculational workload.

Further, an attempt is made through this article 
to generalize SSPT and shown that the new 
formalism unifies SSPT- and LPT-based perturbation 
theories, which resolves difficulties and 
inefficiecies calculating in particular excited 
state corrections without having used tedious 
procedures. The power and elegancy of the unified 
model which is, in a sense, complete are illustrated 
via specifally chosen four examples. We now have 
clear and explicit ways to get corrections to all 
energy levels and state wave functions for a 
given perturbed potential.   

As a concluding remark, SSQM has so far 
illuminated so many different areas of quantum 
mechanics that one might wonder what light 
perturbation theories based on supersymmetry 
might shed on the conventional techniques. In 
that hope, we are not about to be disappointed. 
We believe that this simple and intuitive method 
discussed through this paper would find a widespread 
application in the related area. The application 
of the method to other potentials involving 
Yukawa and Woods-Saxon potentials is in progress.

\end{document}